\documentclass[fleqn,10pt]{wlscirep}
\usepackage[utf8]{inputenc}
\usepackage[T1]{fontenc}
\usepackage{subcaption}
\usepackage{hyperref}
\usepackage{caption}
\usepackage{cleveref}
\usepackage{multirow} 
\usepackage{booktabs}
\title{Seeding hESCs to achieve optimal colony clonality}

\author[1,*]{L E Wadkin}
\author[1]{S Orozco-Fuentes}
\author[2,3]{I Neganova}
\author[2]{S Bojic}
\author[4]{A Laude}
\author[2]{M Lako}
\author[1]{N G Parker}
\author[1]{A Shukurov}

\affil[1]{School of Mathematics, Statistics and Physics, Newcastle University, UK}
\affil[2]{Institute of Genetic Medicine, Newcastle University, UK}
\affil[3]{Institute of Cytology, RAS; Sankt-Petersburg, Russia}
\affil[4]{Bio-Imaging Unit, Medical School, Newcastle University, UK}
\affil[*]{l.e.wadkin@ncl.ac.uk}


\begin{abstract}
Human embryonic stem cells (hESCs) and induced pluripotent stem cells (iPSCs) have promising clinical applications which often rely on clonally-homogeneous cell populations. To achieve this, cross-contamination and merger of colonies should be avoided. This motivates us to experimentally study and quantitatively model the growth of hESC colonies. The colony population is unexpectedly found to be multi-modal. We associate these sub-populations with different numbers of founding cells, and predict their occurrence by considering the role of cell-cell interactions and cell behaviour on randomly seeded cells. We develop a multi-population stochastic exponential model for the colony population which captures our experimental observations, and apply this to calculate the timescales for colony merges and over which colony size no longer predicts the number of founding cells. These results can be used to achieve the best outcome for homogeneous colony growth from different cell seeding densities.
\end{abstract}

\begin{document}

\flushbottom
\maketitle

\thispagestyle{empty}

\section*{Introduction}

Human embryonic stem cells (hESCs) and induced pluripotent stem cells (iPSCs) have promising clinical applications, including the advancement of cellular therapies, disease modelling and drug development, due to their self renewal potential and the ability to differentiate into specialised cells (pluripotency) \cite{pmid26818440, pmid28533935}. However, continued efforts in understanding the complex behaviour of hESCs and iPSCs are necessary to make their clinical applications a reality.

hESCs form colonies by repeated mitosis in which two genetically identical daughter cells are produced from the division of the mother cell. The proliferation of cells in this way results in colonies of tightly packed cells. The doubling time of stem cells varies and can be affected by various environmental and chemical factors, including cell density \cite{BARBARIC2014142, pmid19235204, pmid25412279}. 

An important measure of the self-renewal potential of stem cells is the clonality, the condition of being genetically identical. Generating homogeneous populations of clonal cells is of great importance \cite{GLAUCHE2013232, HEINS2006511} as clonally derived stem cell lines maintain pluripotency and proliferative potential for prolonged periods \cite{AMIT2000271}. Some applications require clonal homogeneous populations, e.g. drug discovery \cite{pmid21217770} and iPSCs for personalised medicine. The selection of the best clones for further experimentation needs to be optimised to make clinical applications safe. At high seeding densities the migration of cells and the growth of closely-separated cell groups can cause aggregation of colonies; this is undesirable when a homogeneous clonal population with identical genetic composition is required. The seeding density of cells has been shown to not only have an effect on the clonality of stem cells \cite{Li}, but also on their differentiation potential \cite{pmid24324748}. Moreover, culturing at an overly high density can cause DNA damage and culture adaptation, leading to increasing occurrence of chromosomal aberrations \cite{pmid26923824, BARBARIC2014142, andrews}.

Single hESCs are reported to have no effect on each other's movement if they are greater than 150\,$\mu$m apart \cite{Li}. It is therefore recommended to keep a minimum distance of 150\,$\mu$m between colony boundaries throughout growth to assure the resulting clonal structures are from single founding hESCs. In our previous work we considered the kinematics of single and pairs of cells, and identified the occasional super-diffusive movements of cells which could lead to re-aggregation \cite{me1, me2}. 

Here we focus on quantifying the possibilities of re-aggregration due to the physical proximity of colonies at different seeding densities and consider the optimal seeding densities to form clonal structures. From experimental observations of colonies after a fixed evolution time, we infer a stochastic model for colony growth, a method previously applied to other cell types, including bacterial and cancerous cells \cite{PhysRevLett.113.028101,pmid29201315}. To correctly initialize the model for a given seeding density, we take into account the proportion of seeded cells which begin as isolated cells, pairs, triples, and so on.  We find that this is essential to capture the experimental observations. We use the model to simulate hESC colony growth at different seeding densities, consider the area coverage with time and calculate the critical time at which the homogeneous colonies begin to merge. These results can help inform cell seeding decisions to form homogeneous colonies from single founder cells.

\section*{Results}

\paragraph{Notation} The notation used throughout the manuscript is outlined here for convenience. The number of cells in a colony at a given time, $t$, is $N(t)$, with $t$ always in hours. Therefore the number of cells at 72\,h is $N(72)$. The initial number of cells at time $t=0$ is $N(0)\equiv N_0$. The seeding density is $n_0$ and the density of attached cells after 24\,h is $\eta_0=0.35n_0$, given in cells/cm$^2$. The growth rate of a colony is $\gamma$ (given in $\textrm{h}^{-1}$), the division rate is 1/$\gamma\,$ (given in h) and the population doubling time is $t_{\textrm{d}}$ (given in h). The time at which the number of founder cells is indistinguishable based on colony size is $t_\ast$ (given in h). The average time at which colonies merge due to physical proximity is $\tau$ (given in h).

\subsection*{Experimental colony size}

\begin{figure}[ht]
\centering
\begin{subfigure}[t]{0.47\textwidth}
\caption{}
\includegraphics[width=1\textwidth]{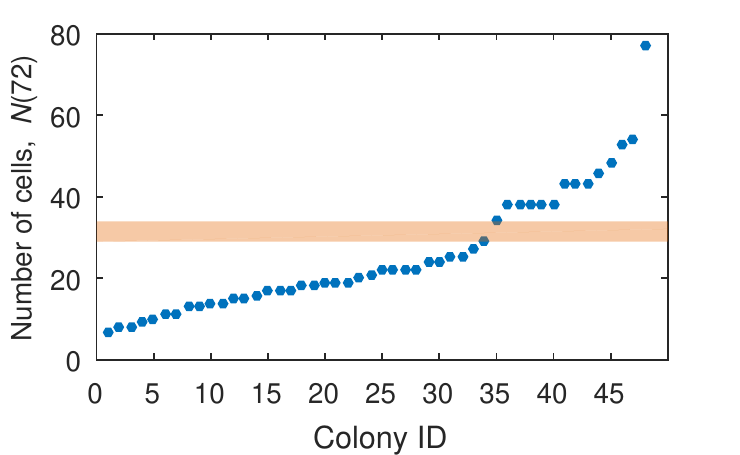} 
\label{fig:N72a}
\end{subfigure}
\begin{subfigure}[t]{0.475\textwidth}
\caption{}
\includegraphics[width=1\textwidth]{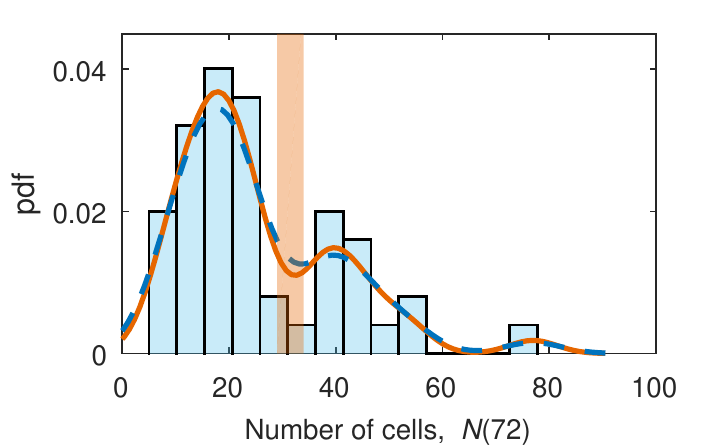} 
\label{fig:N72b}
\end{subfigure}
\\
\begin{subfigure}[t]{0.45\textwidth}
\caption{}
\vspace{10pt}
\includegraphics[width=1\textwidth]{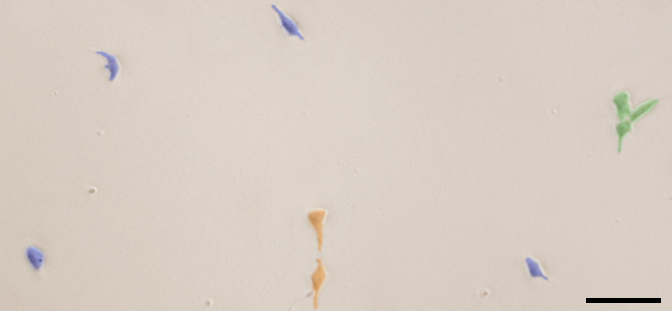}
\label{fig:seedingschematic}
\end{subfigure}
\hspace{10pt}
\begin{subfigure}[t]{0.49\textwidth}
\centering
\caption{}
\includegraphics[width=1\textwidth]{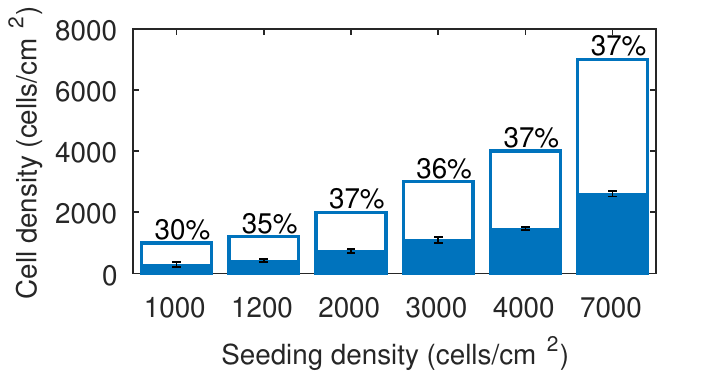}
\label{fig:attachment} 
\end{subfigure}
\caption{\label{fig:fig1}(\subref{fig:N72a}) $N(72)$ for each colony arranged in ascending order. The horizontal orange block shows the splitting of the data into two groups using K-means clustering. The split occurs between $29\leq N(72)\leq34$.  (\subref{fig:N72b}) A histogram of $N(72)$ with kernel density estimates with bandwidths of 4.5 (orange) and 5.43 (auto bandwidth, blue dashed). The vertical orange block shows the splitting of the data into the two groups using K-means clustering. (\subref{fig:seedingschematic}) Microscopy image of cells seeded at density $n_0=1200\,$cells/cm$^2$. Most cells have no neighbours within the critical interaction distance of 150\,$\mu$m (highlighted in blue), but some are in pairs (orange) and triples (green). The scale bar represents 100\,$\mu$m. (\subref{fig:attachment}) Bar chart showing the initial cell seeding density (cells/cm$^2$) at day zero ($n_0$, unfilled bar height), and the mean cell density of the remaining cells attached to the matrix at day one ($\eta_0$, filled bars). Each mean is based on three independent measurements and the error bars represent standard deviations.}
\end{figure}

From Experiment 1, the number of cells in each of the 48 colonies at 72 hours after cell attachment, $N$(72), was recorded and is shown for each colony arranged in ascending order in Figure~\ref{fig:N72a}. The corresponding histogram of $N$(72) is shown in Figure~\ref{fig:N72b}. The distribution is bimodal, confirmed by the kernel density estimation, with an outlier colony at $N$(72)$=77$ cells. We remove this outlier colony for further analysis as it was most likely formed by several colony merges. We expect the number of cells at time $t$ to evolve roughly as $N(t)=N_02^{t/t_\mathrm{d}}$, where $N_0\equiv N$(0) is the initial number of cells and $t_{\mathrm{d}}$ is the time it takes for the population to double, or equivalently $N(t)=N_0e^{\gamma t}$, where $\gamma$ is the growth rate. The bimodal nature of $N(72)$ implies that we have two distinct groups of colonies, lead by differences in $N_0$ and/or differences in the growth rates between colonies. For a typical duration of the cell cycle, 16--18\,hours, one expects 20 cells at 72 hours, corresponding to the first histogram peak. The second peak, at about 40 cells per colony, suggests that some groups of cells have merged to form larger colonies during the 72 hour period, or that the initial condition of the colony growth was in fact $N_0=2$. K-means clustering, a standard algorithm which partitions observations into clusters based upon minimising within cluster variance, splits $N$(72) into the two ranges, $7\leq N(72)\leq 29$ and $34\leq N(72)\leq 77$.

To ascertain the initial conditions that underlie the colony growth, we turn to Experiment 2, examining the cells after 24 hours; a typical image is shown in Figure~\ref{fig:seedingschematic}. There are several characteristic features of the cell distribution revealed in this experiment. Firstly, the random initial positioning of the seeded cells means that some cells are initially isolated with no cells within the interaction distance (estimated as 150$\,\mu$m). Other cells lie within the interaction distance of each other, forming groups of varying sizes.  In Figure~\ref{fig:seedingschematic} we colour-code the cells according to whether they are isolated, or are effectively in a pair or in a triple, to illustrate how $N_0$ can vary at low seeding densities. We will return to this feature later.

Secondly, only a fraction of the originally seeded cells are attached to the plate at this time.  We find for a range of seeding densities, $1000\leq n_0 \leq7000\,$cells/cm$^2$, that on average $35.19\%\pm4.23\,[0.99]$ (the mean $\pm$ one standard deviation [standard error]) of initially seeded cells were attached 24 hours after plating. Figure~\ref{fig:attachment} shows the proportion of attached cells at different seeding densities. In the following modelling section of this paper we discuss $N(t)$, the number of cells present in a colony over time, independent of original cell seeding densities. In the cell seeding section we discuss the effects of the cell plating densities, $n_0$, where we assume that the actual density of cells present is $\eta_0=0.35n_0$ to account for the loss at the attachment stage.

\newpage
\subsection*{Development of the exponential growth model}\label{sec:modelling}

Throughout this paper we define time $t=0$ to be the time that seeded cells have attached to the plate and their proliferation starts. Before this time some cells are lost (as reported above and shown in Figure~\ref{fig:attachment}) and there is a delay in the growth from the lag-phase experienced by newly plated \textit{in vitro} hESCs as they adjust to the environment \cite{pmid21822870}. This is consistent with the experimental data which considers 72\,h after cell attachment. 

The simplest deterministic model for the number of cells in a colony at time $t$, $N(t)$, assumes a constant cell division time $1/\gamma\,$ and simultaneous division of all the cells, leading to 

\begin{equation*}
\frac{dN(t)}{dt}=\gamma N,
\label{eq:detN}
\end{equation*}

\noindent
which has the solution, 

\begin{equation*}
N(t)=N_0e^{\gamma t},
\end{equation*}

\noindent
where $N_0\equiv N(0)$ is the initial number of cells at $t=0$. However, the cell cycle duration is variable due to various factors, such as inhomogeneities in the nutrient distribution within the growth medium and the inherent variation in the cell cycle duration between different clones. Such effects can be allowed for by considering a Gaussian random growth rate $\gamma$, with a mean value $\mu_0$ and standard deviation $\sigma_0$:

\begin{equation}
N(t)=N_0e^{\gamma t}, \,\,\, \gamma\sim \rm{Norm}(\mu,{\sigma}^2).
\label{eq:model}
\end{equation}
 
\noindent
Different colonies thus grow at different rates sampled from the Gaussian probability distribution. The number of cells then follows a lognormal distribution, $N(t) \sim\rm{LogNorm}(\mu_0,{\sigma_0}^2)$, where $\mu_0=t\mu+\log(N_0)$ and $\sigma_0=t^2{\sigma}^2$. A short mathematical explanation can be found in the Supplementary Information. However, this model fails to explain the bimodal distribution of the colony sizes observed at $t=72\,$h. This is presented in the Supplementary Information, Figure S1.

We suggest that a bimodal distribution of the colony sizes can be a consequence of a difference in the cell proliferation rates in cell groups of different sizes that may arise from their interactions. It can be expected that colonies starting from larger groups grow faster. To capture the bimodal nature of the colony size distribution, we consider two populations, A and B, each with a different initial condition,

\begin{equation*}
    N_0=
    \begin{cases}
      1, & \text{group A},\\
      2, & \text{group B},\\
    \end{cases}
  \end{equation*}

\noindent
where the probabilities for a colony to belong to groups A and B are $\alpha$ and $\beta$, respectively. Each population then follows equation ~(\ref{eq:model}) with its corresponding initial condition,

\begin{equation}
\begin{cases}
N_\textrm{A}(t)=e^{\gamma t}, \,\,\, \gamma\sim \rm{Norm}(\mu_{A}, {\sigma_{A}}^2), & \text{with probability $\alpha$},\\

N_\textrm{B}(t)=2e^{\gamma t}, \,\,\, \gamma\sim \rm{Norm}(\mu_{B}, {\sigma_{B}}^2), & \text{with probability $\beta$,}
\end{cases}
\label{eq:model2}
\end{equation}

\noindent
and each of $N_\textrm{A}$ and $N_\textrm{B}$ has a lognormal probability distribution. Thus we consider separately colonies that originate from a single cell and those from cell pairs. We consider the possible role of cell triples and larger progenitor groups in the Discussion.

\begin{figure}[ht]
\begin{subfigure}[t]{0.49\textwidth}
\caption{}
\includegraphics[width=1\textwidth]{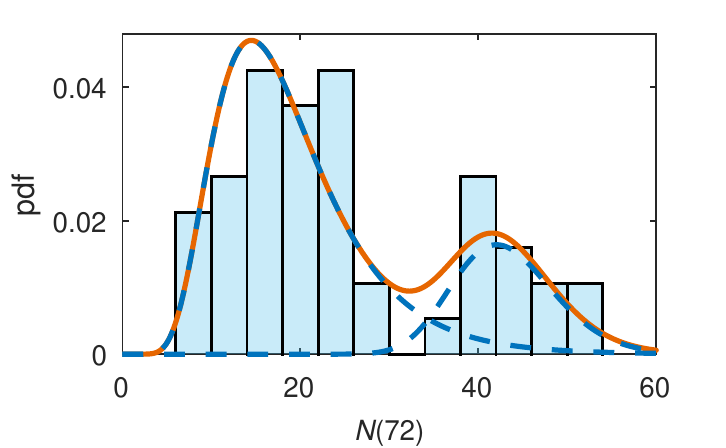}
\label{fig:lognormalmix}
\end{subfigure}
\begin{subfigure}[t]{0.49\textwidth}
\caption{}
\includegraphics[width=1\textwidth]{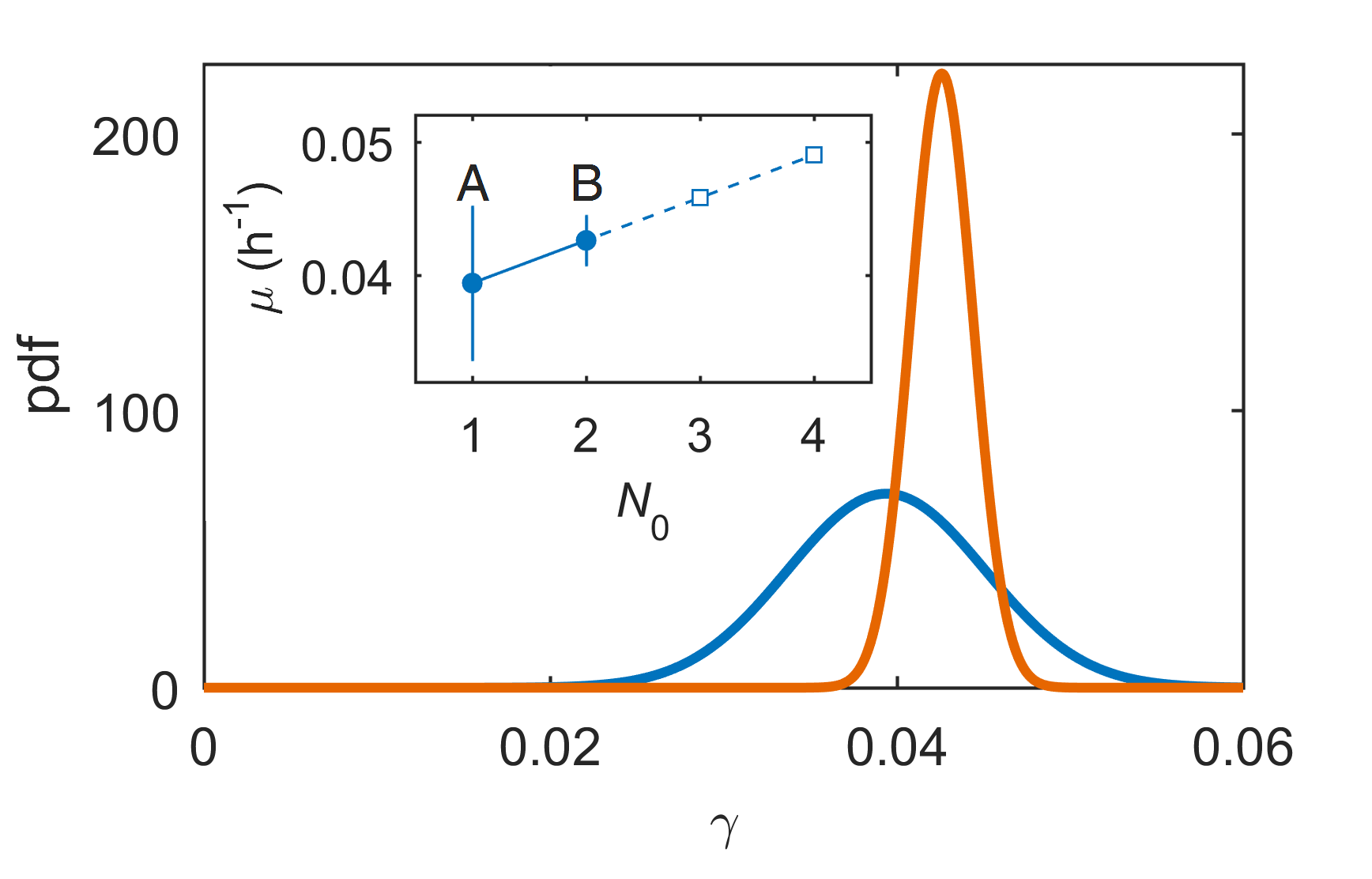}
\label{fig:gammadist}
\end{subfigure}
\caption{\label{fig:lognormalmixture} (\subref{fig:lognormalmix}) $N(72)$ with a lognormal mixture model fitting corresponding to the two populations. The population fittings, $N_\textrm{A}(72)\sim \rm{LogNorm}(2.84,0.41^2)$ with mixture probability 0.77 and $N_\textrm{B}(72)\sim \rm{LogNorm}(3.76,0.13^2)$ with mixture probability 0.23 are shown in blue dashed, and the overall mixture distribution in orange. (\subref{fig:gammadist}) The two $\gamma$ distributions with group A in blue and group B in orange. Note that the distributions are not scaled to represent the group probabilities $\alpha$ and $\beta$. The inset shows the parameters for $\gamma$, $\mu$ with $\,\pm\,\sigma$ error bars for the initial conditions $N_0=1$ and $N_0=2$, corresponding to group A and group B, respectively. The dashed line shows the extrapolation of the trend to higher values of $N_0$.}
\end{figure}

A lognormal mixture fit to the data, shown in Figure~\ref{fig:lognormalmix}, gives $N_\textrm{A}(72)\sim\rm{LogNorm}(2.84,0.41^2)$ and $N_\textrm{B}(72)\sim\rm{LogNorm}(3.76,0.13^2)$, with the mixture probabilities $\alpha=0.77$ and $\beta=0.23$. Therefore, we have $\mu_{A}=0.0394$\,h$^{-1}$, $\sigma_{A}=0.0057$\,h$^{-1}$, $\mu_{\textrm{B}}=0.0426$\,h$^{-1}$, and $\sigma_{\textrm{B}}=0.0018$\,h$^{-1}$. A comparison between these growth parameter distributions is shown in Figure~\ref{fig:gammadist}. These values correspond to a doubling time of 17.5\,h for the single founder cell population (group A) and 16.3\,h for the pairs of cells population (group B).

\subsubsection*{Modelling population growth}

\begin{figure}[!]
\centering
\begin{subfigure}{0.49\textwidth}
\caption{}
\includegraphics[width=\textwidth]{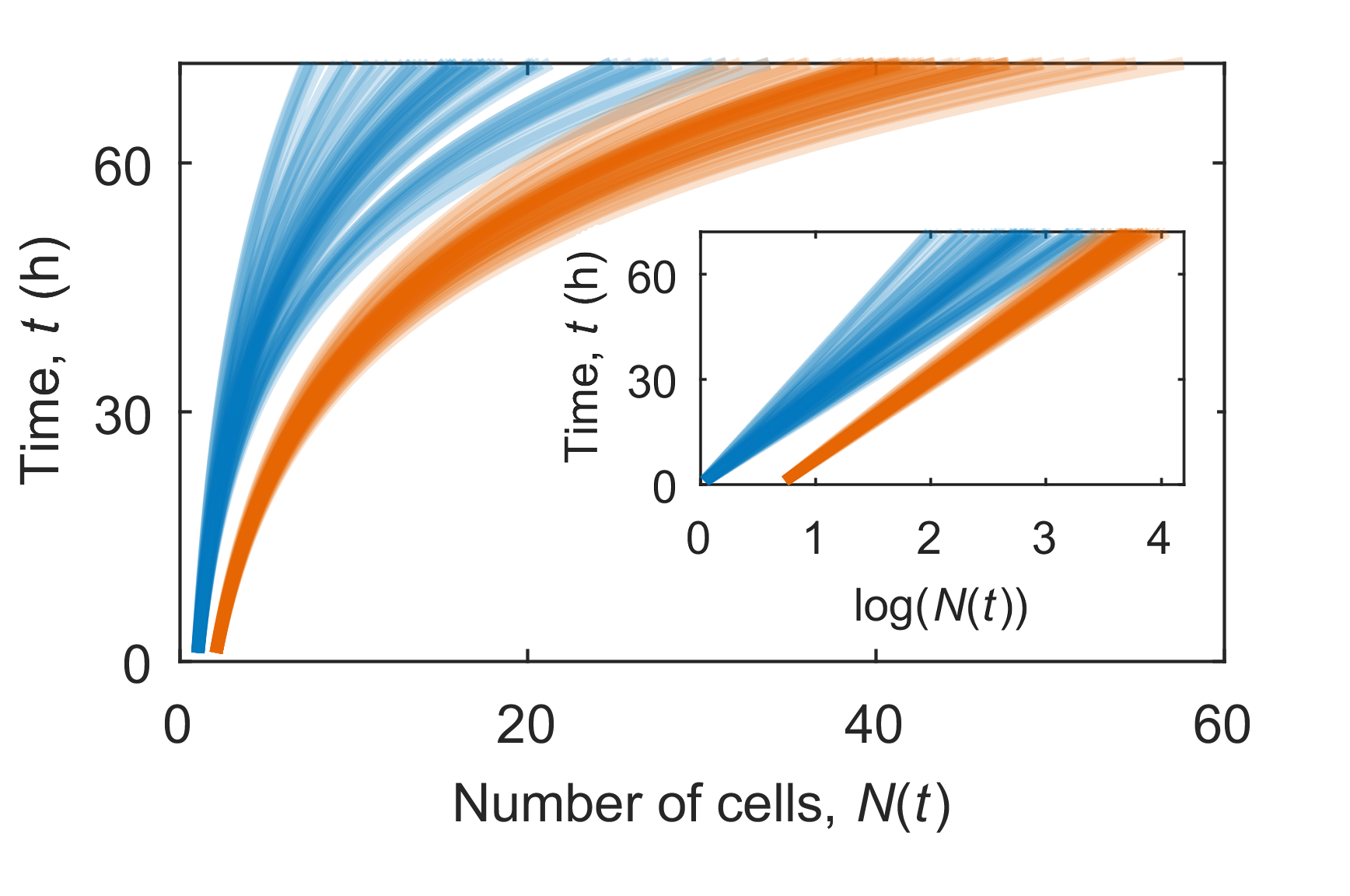}
\label{fig:N50}
\end{subfigure}
\begin{subfigure}{0.325\textwidth}
\caption{}
\includegraphics[width=\textwidth]{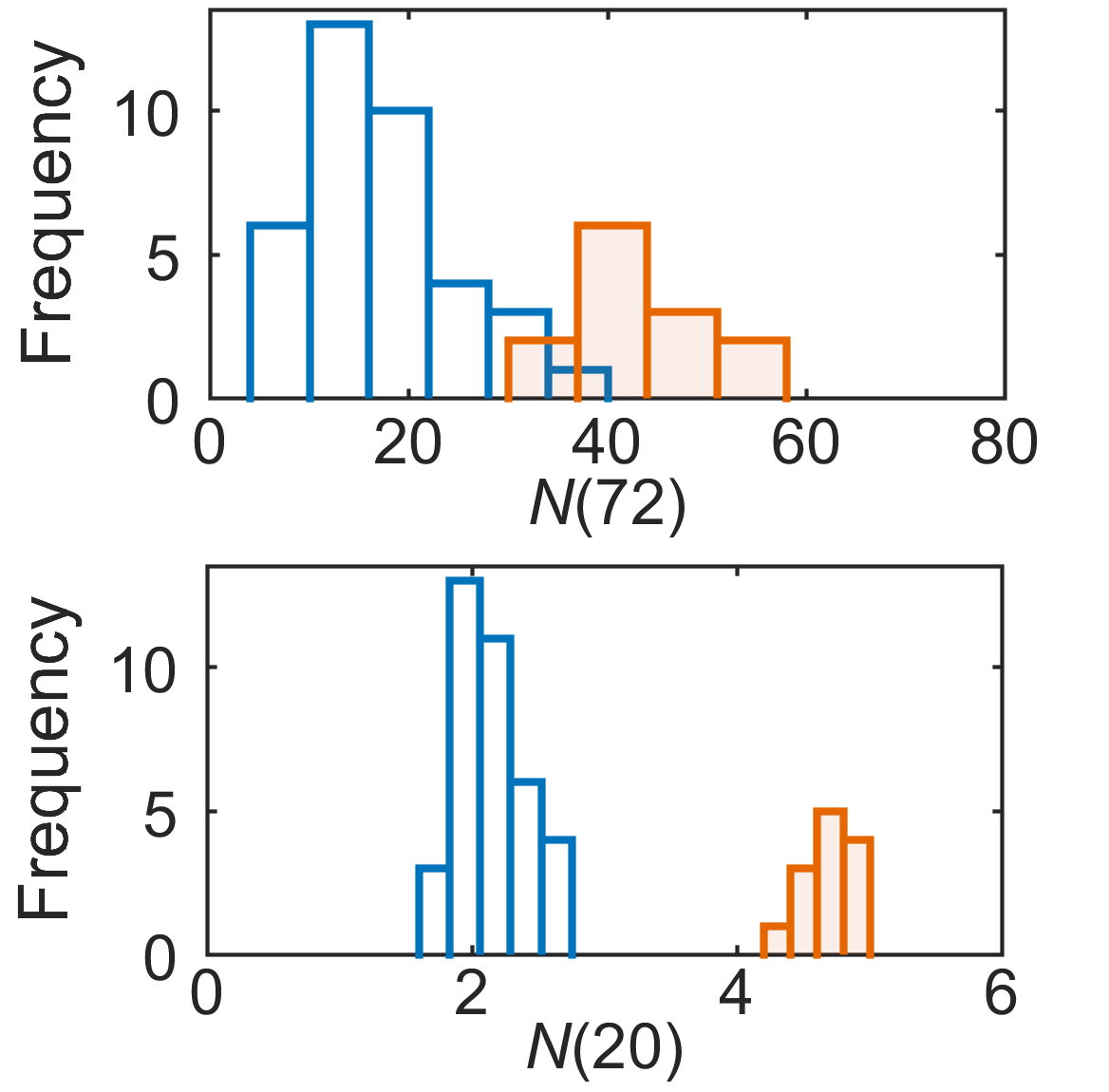}
\label{fig:Nhist1}
\end{subfigure}
\\
\begin{subfigure}{0.49\textwidth}
\caption{}
\includegraphics[width=\textwidth]{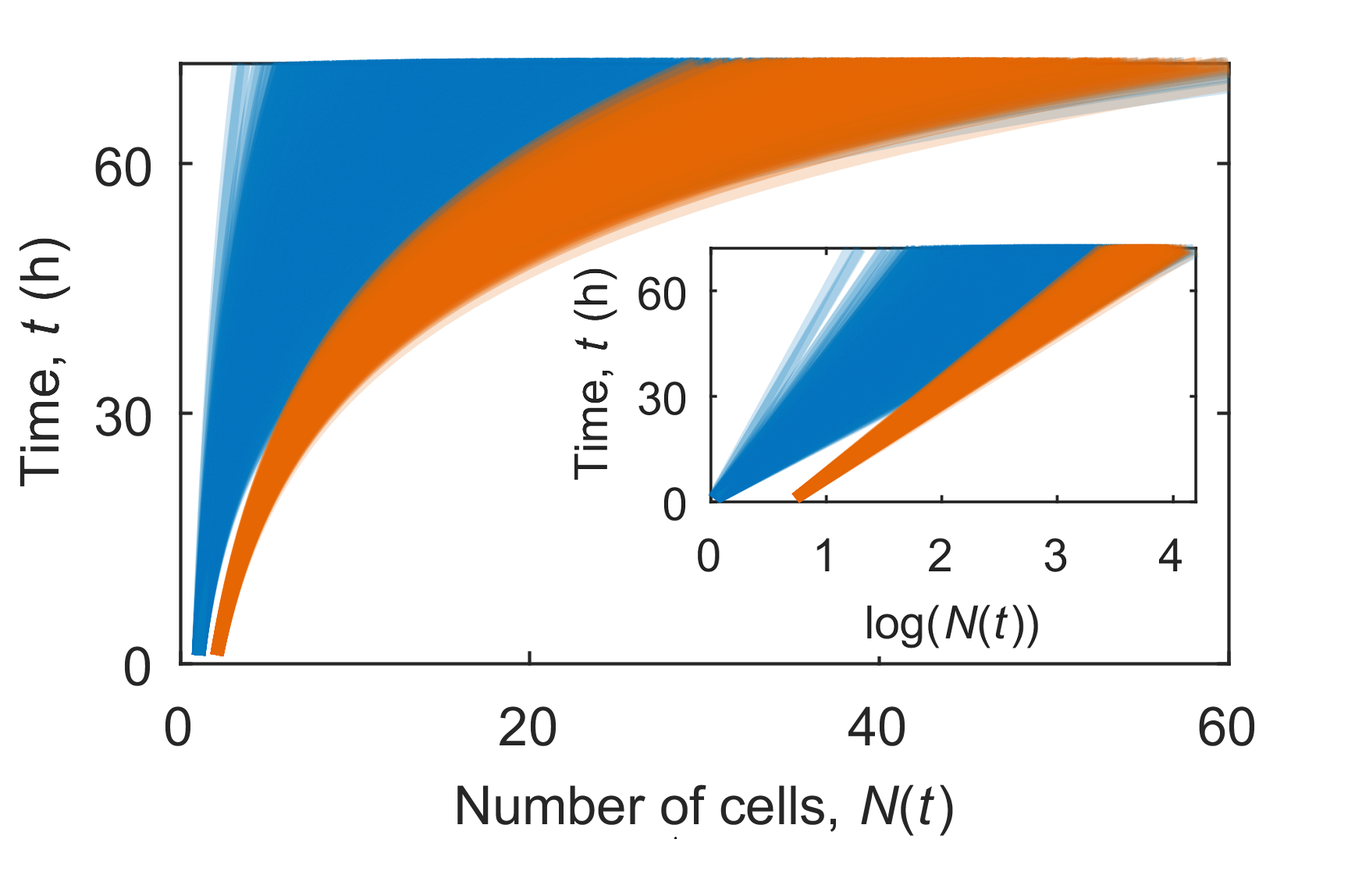}
\label{fig:N5000}
\end{subfigure}
\begin{subfigure}{0.325\textwidth}
\caption{}
\includegraphics[width=\textwidth]{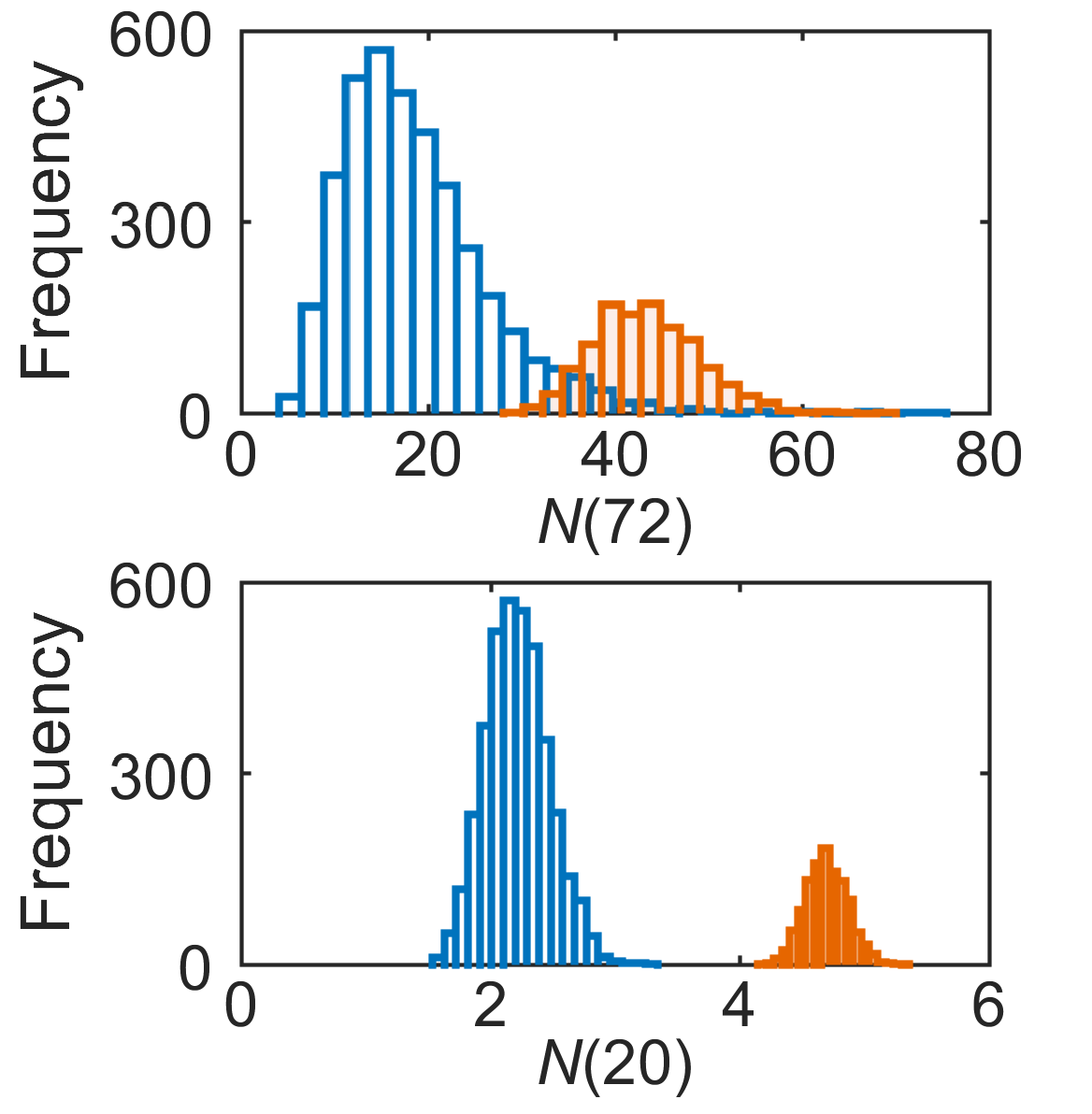}
\label{fig:Nhist3}
\end{subfigure}
\caption{\label{fig:Nevo} Evolution of $N(t)$ from the two-population model for (\subref{fig:N50}) 50 and (\subref{fig:N5000}) 5000 colonies, with blue $N_{\textrm{A}}(t)$ ($N_0=1$) and orange $N_{\textrm{B}}(t)$ ($N_0=2$). The insets show the corresponding plots of log($N(t)$). Histograms of $N(t)$ for (\subref{fig:Nhist1}) 50 and (\subref{fig:Nhist3}) 5000 colonies at $t=20\,$h and $t=72\,$h.}
\end{figure}

\begin{figure}[!]
\centering
\includegraphics[width=0.49\textwidth]{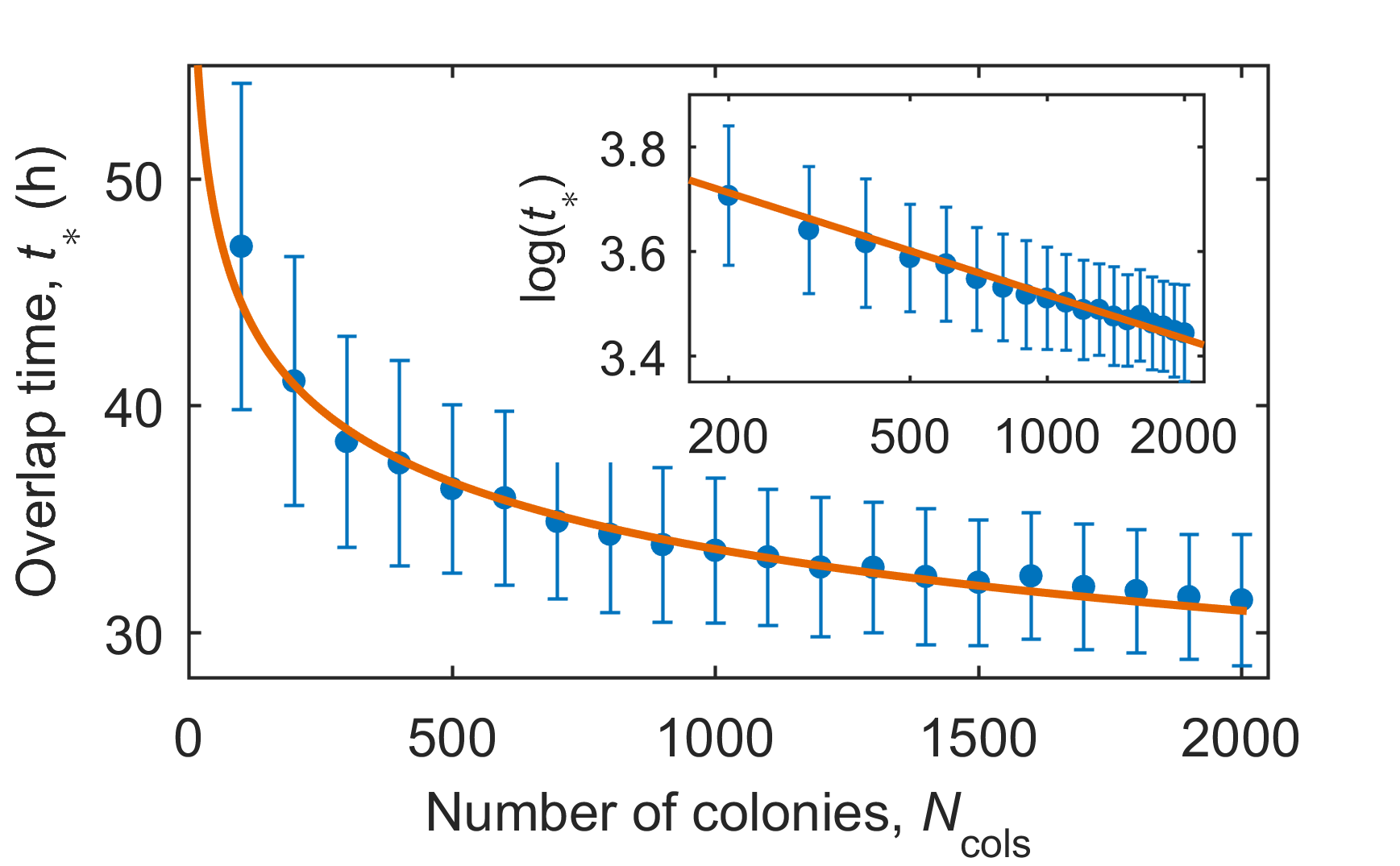}
\label{fig:logoverlaptimes}
\caption{\label{fig:overlaptimeall}The mean critical time, $t_{\ast}$ at which colonies originating from a single cell are no longer identifiable based on colony size, $N(t)$. Each data point is the mean of 500 simulations. The error bars represent the standard deviations in the means. The data is well captured by a power law relation $t_{\ast}=a{N_{\rm{cols}}}^{-b}$ (orange line, $R^2=0.98$); linear least-squares fitting on a log-log plot (inset) gives $a=77.9\pm4.7\,$h and $b=0.12\pm0.01$.}
\end{figure}

We have demonstrated how accurately the two-population model captures the experimental data at 72 hours. Now we proceed to develop it into a prognostic model for the colony size at later times. The evolution of the colony size, $N(t)$, according to this two-population model is shown in Figure~\ref{fig:Nevo}. Because of the random scatter in the colony growth parameters the admissible range of colony size $N(t)$ increases as $\sqrt{t}$, and sooner or later, the size of the two colony types overlap. At early times, the sizes of the two colony types are distinct, where those beginning from two cells are larger than those from one founder cell, but as time progresses the stochasticity in the growth rates causes an overlap in the two populations. This overlap becomes more prominent for larger numbers of simulations as this increases the incidence of extreme growth rate values. Histograms of $N(20)$ and $N(72)$ in Figures~\ref{fig:Nhist1} and \ref{fig:Nhist3} illustrate how at early times the two colony types are distinguishable but over time the distributions spread and merge to make single-clone colonies indistinguishable from heterogeneous colonies. The time at which $N_{\textrm{A}}$ first becomes equal to $N_{\textrm{B}}$, $t_\ast$, is the critical time after which it is not possible to distinguish which colonies originated from a single progenitor based on the colony size. This time is shown for increasing numbers of colonies in Figure~\ref{fig:overlaptimeall}. As we increase the number of colonies, $N_{\textrm{cols}}$, we see more of the extreme values occurring with low probability, causing the blurring of the two populations to begin at an earlier time. The relationship is a power law, with the best fit ${t_{\ast}}=a{N_{\rm{cols}}}^{-b}$ with $a=77.9\pm4.7\,$h and $b=0.12\pm0.01$ with an $R^2$ coefficient of 0.98. This allows us to estimate, for a given plating cell density, the time up to which colonies originating from a single founder cell are identifiable based on the current number of cells in the colony. 

\subsection*{Role of seeding density and cell clustering on the formation of homogeneous hESC colonies}\label{ssec:seeding}

Typical low seeding densities for hESCs, intended to grow colonies from single founder cells, commonly range from 500 to 3000\,cells/cm$^2$. Across a range of seeding densities, we find that the average proportion of cells attached to the substrate after 24 hours is $35\pm4\%$, where the range represents one standard deviation within the sample, and the accuracy of the mean value (the standard error) is $\pm 1.0\%$, presented in Figure~\ref{fig:attachment}. For example, an initial seeding density of $n_0=1500\,$ cells/cm$^2$ results in around 500 cells continuing past day one of the experiment. Throughout this section we will present the initial seeding densities $n_0$ and work on the assumption that 35\% of these cells are successfully attached and survive, $\eta_0=0.35n_0$.

In this section we find the initial conditions corresponding to cell plating at different cell seeding densities and use this to inform the model for colony growth. The seeding of cells randomly across a growth area, $A$, can be simulated as a homogeneous Poisson point process in which the number of point counts is sampled from a Poisson distribution with the mean $\lambda A$. The points are then independently and uniformly scattered across the region. For example, if we consider an initial seeding density of $n_0=1500$\,cells/cm$^2$, we can simulate the seeding of $\eta_0=0.35n_0$ cells in a $1\,$cm$^2$ area by a homogeneous point Poisson process in which the $\eta_0$ point counts are sampled from Po($\lambda A$) where $A=1$\,cm$^2$, and then locating cells according to the uniform distributions in $x$ and $y$. 

\begin{figure}[!]
\centering
\begin{subfigure}{0.49\textwidth}
\caption{}
\includegraphics[width=\textwidth]{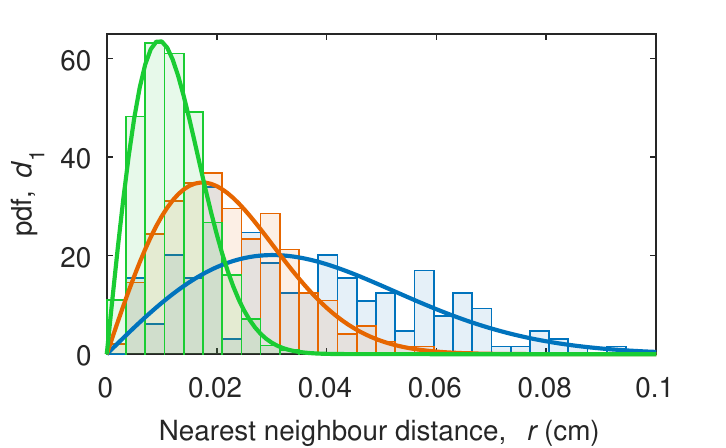}
\label{fig:nnhists}
\end{subfigure}
\begin{subfigure}{0.49\textwidth}
\caption{}
\includegraphics[width=\textwidth]{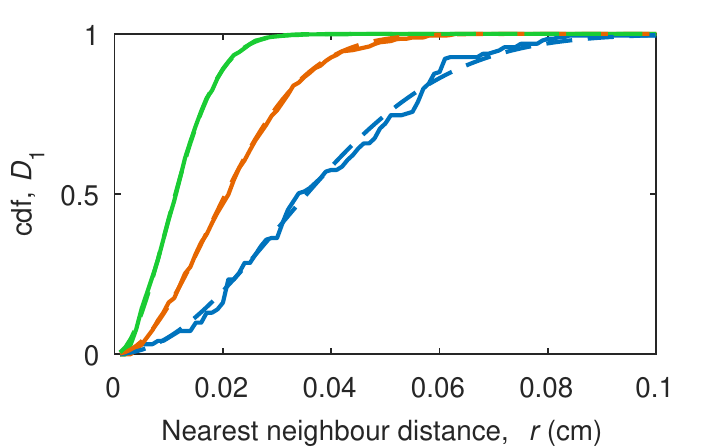}
\label{fig:nncdf}
\end{subfigure}
\\
\hspace{10pt}
\begin{subfigure}{0.4\textwidth}
\caption{}
\includegraphics[width=\textwidth]{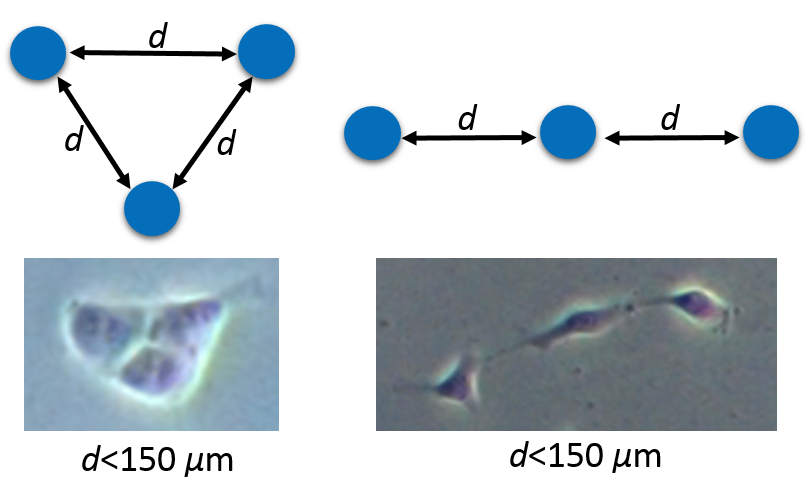}
\label{fig:clusteringdiag}
\end{subfigure}
\hspace{20pt}
\begin{subfigure}{0.45\textwidth}
\caption{}
\includegraphics[width=\textwidth]{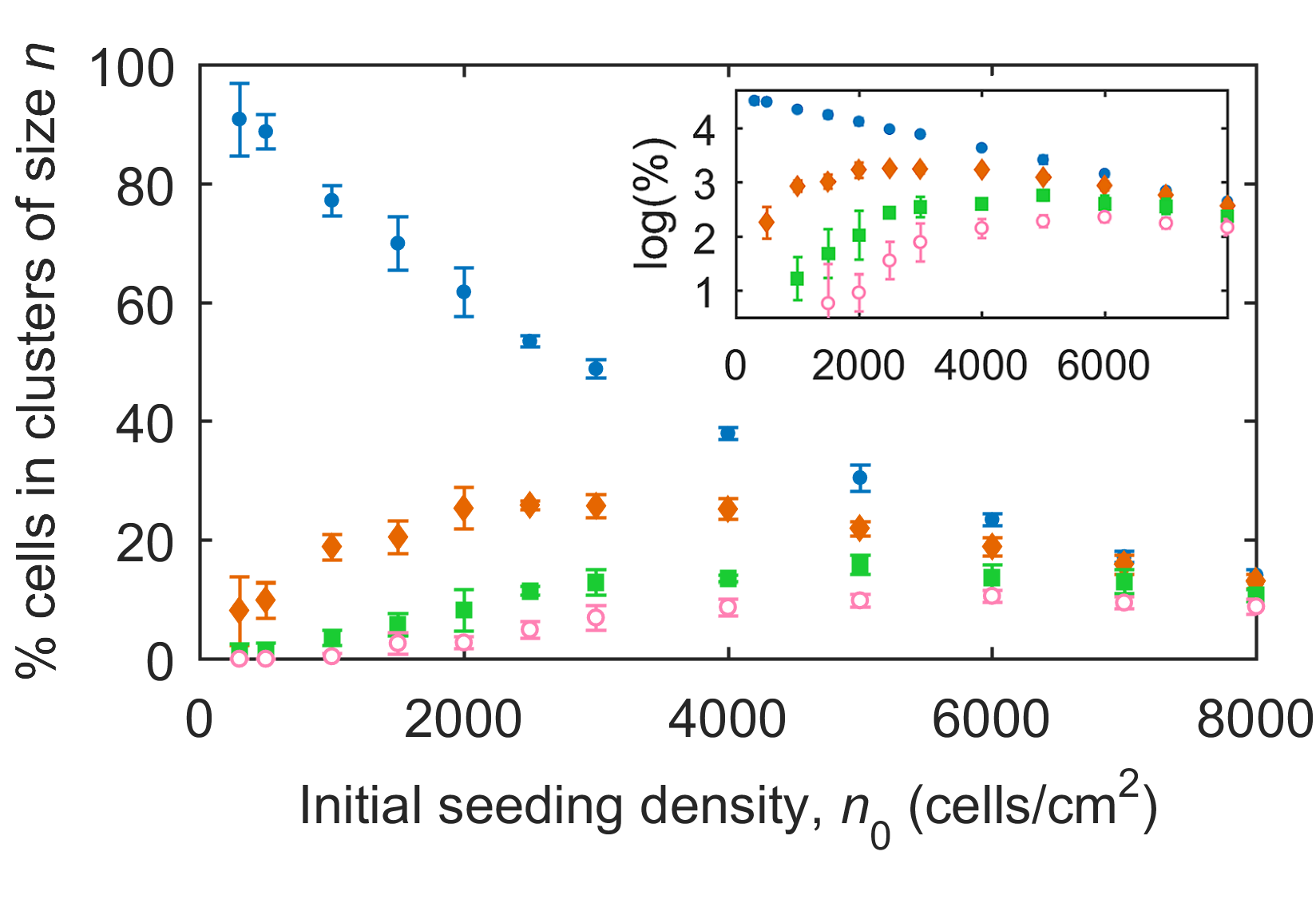}
\label{fig:clusteringgraph}
\end{subfigure}
\caption{\label{fig:nearestn}(\subref{fig:nnhists}) The probability distributions for the nearest neighbour distance in a Poisson point process, given by $d_1(r)=2\lambda\pi r e^{-\lambda \pi r^2}$ for $\lambda=\eta_0=175$ (blue, $n_0=500$\,cells/cm$^2$), $\lambda=\eta_0=525$ (orange, $n_0=1500$\,cells/cm$^2$) and  $\lambda=\eta_0=1750$ (green, $n_0=5000$\,cells/cm$^2$). The histograms show the simulated distributions for seeding cells according to a Poisson process with rate $\lambda$, each is for one run of the seeding simulation. (\subref{fig:nncdf}) The theoretical proportion of cells with a nearest neighbour $<d\,$cm away, $D_{1}(r)=1-e^{-\lambda\pi r^2}$, for $\lambda=\eta_0=175$ (blue dashed, $n_0=500$\,cells/cm$^2$), $\lambda=\eta_0=525$ (orange dashed, $n_0=1500$\,cells/cm$^2$) and $\lambda=\eta_0=1750$ (green dashed, $n_0=5000$\,cells/cm$^2$). The corresponding solid lines show the values of $D_1$ from a single run of the seeding simulation. (\subref{fig:clusteringdiag}) Example clustering configurations for cell triples with microscopy examples of these formations. The distance between cells, $d$, must be less than 150\,$\mu$m. (\subref{fig:clusteringgraph}) The percentage of cells in a cluster of size $n$ (for an interaction radius of $150\,\mu$m) for $n=1$ (single cells, blue circles), $n=2$ (pairs, orange diamonds), $n=3$ (triplets, green square) and $n=4$ (quadruples, pink open circles). The error bars represent standard deviations. Inset: The data on a log-log scale. The convergence to a linear relationship (and hence an exponential relationship for the non transformed data) can be seen.}
\end{figure}

Once cells have been scattered, we can then consider the distances between cells and their nearest neighbours with the aim to estimate the fraction of isolated cells, their pairs (defined as two cells separated by less than 150\,$\mu$m) and triples etc. The probability density function of the distance, $r$, to the $k^{\rm{th}}$ nearest neighbour is known from the theory of Poisson point processes as $d_k(r)=2(\lambda\pi r^2)^k e^{-\lambda \pi r^2}/r(k-1)!$ \cite{spatstat}. This reduces to $d_1(r)=2\pi\lambda r^2e^{-\lambda\pi r^2}$ for the first nearest neighbour. The theoretical distributions along with histograms from simulated data for $d_1(r)$ are shown in Figure~\ref{fig:nnhists} for initial seeding densities of $n_0=500$, 1500 and 5000\,cells/cm$^2$ corresponding to $\lambda=\eta_0=0.35n_0$. These distributions allow us to calculate the proportion of seeded cells with the nearest neighbour at a given distance. The nearest neighbour cumulative distribution function for the proportion of cells with a nearest neighbour at a distance $<r$ for a 2D homogeneous Poisson process is given by $D_{1}(r)=1-e^{-\lambda\pi r^2}$ \cite{spatstat}. This theoretical proportion of cells with a first nearest neighbour less than $r$ away, $D_1$, for initial seeding densities $n_0=500$, 1500 and 5000\,cells/cm$^2$ is shown in Figure~\ref{fig:nncdf} along with data from a simulation at each seeding density. For the initial seeding density of $n_0=1500\,$cells/cm$^2$ the nearest neighbour distance between cells will be less than $150\,\mu$m in around $30\%$ of cases, similar to the experimental estimate of 23\%. We have neglected the movement of cells as, based on observed migration speeds of approximately $16\,\mu$m/h \cite{me1}, the time required to traverse the critical interaction distance of $150\,\mu$m is around 9\,h, a large portion of the cell cycle time. 

To consider the groupings of seeded cells we use a density based clustering algorithm. Cells less than 150$\,\mu$m apart are considered as being part of the same cluster, and any neighbouring cell less than 150$\,\mu$m away from any other cell in the cluster is also considered to be part of the cluster. This allows for clusters of elongated shapes. Note that this definition of a cluster of cells is non-trivial and has implications for the interactions of cells, but from the experimental images we see examples of clusters in elongated shapes as well as the more common regular shapes. Examples of different cluster formations for triples are shown in Figure~\ref{fig:clusteringdiag}. The proportion of cells in a cluster of a size $n$ at different seeding densities is shown in Figure~\ref{fig:clusteringgraph}.  At low initial seeding densities, e.g. $n_{\rm{0}}=500$\,cells/cm$^2$ the majority of cells have no close neighbours. As the seeding density increases, the proportion of pairs of cells increases. The proportion of each cluster size first rises with $n_0$ before reaching a maximum and then tends to zero as more possible cluster sizes become available. The distributions shown in Figure~\ref{fig:nearestn} provide the initial conditions corresponding to cell seeding at different densities. 

\begin{figure}[!]
\centering
\begin{subfigure}{0.49\textwidth}
\caption{}
\includegraphics[width=\textwidth]{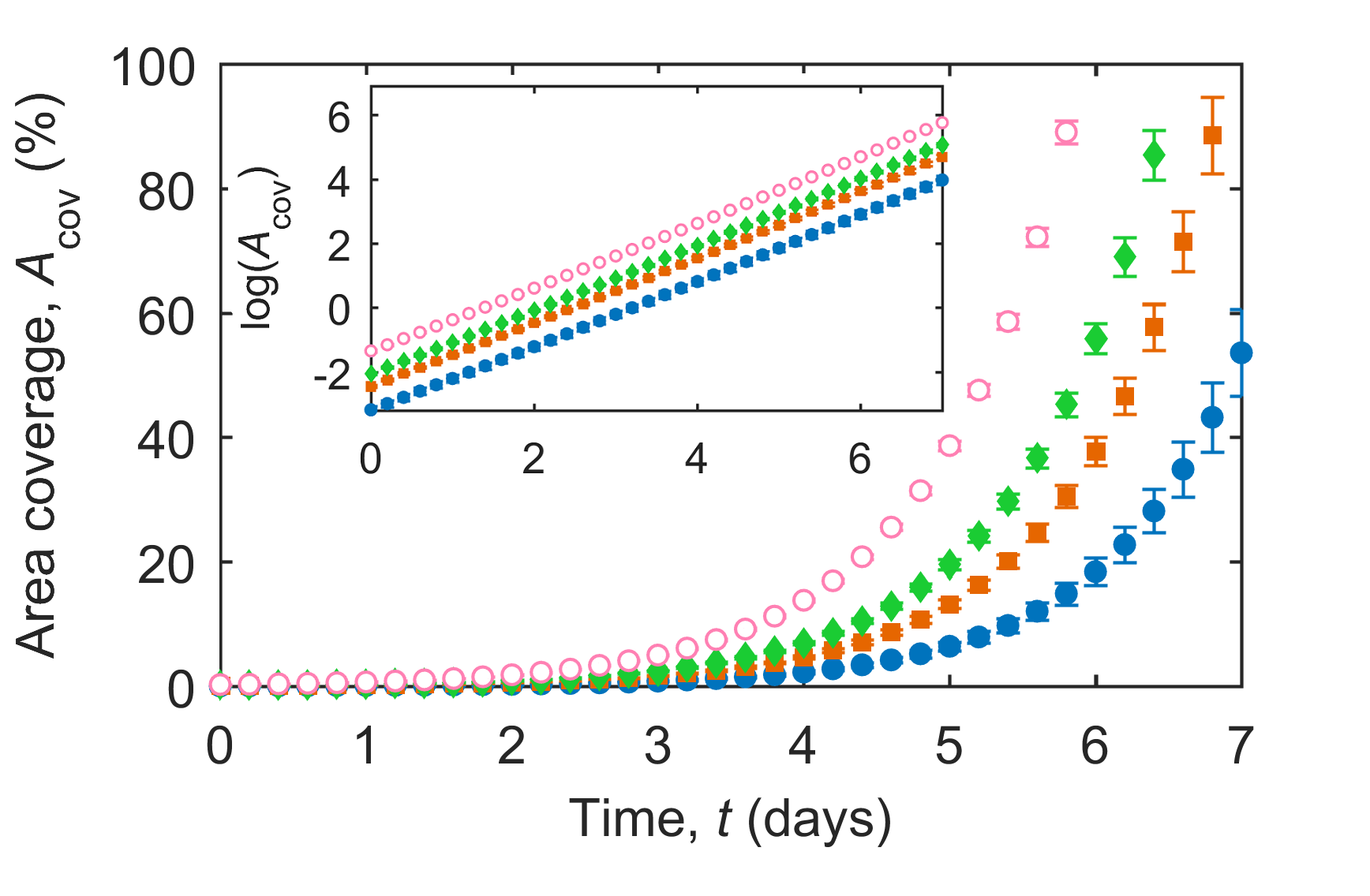}
\label{fig:coverage1}
\end{subfigure}
\begin{subfigure}{0.49\textwidth}
\caption{}
\includegraphics[width=\textwidth]{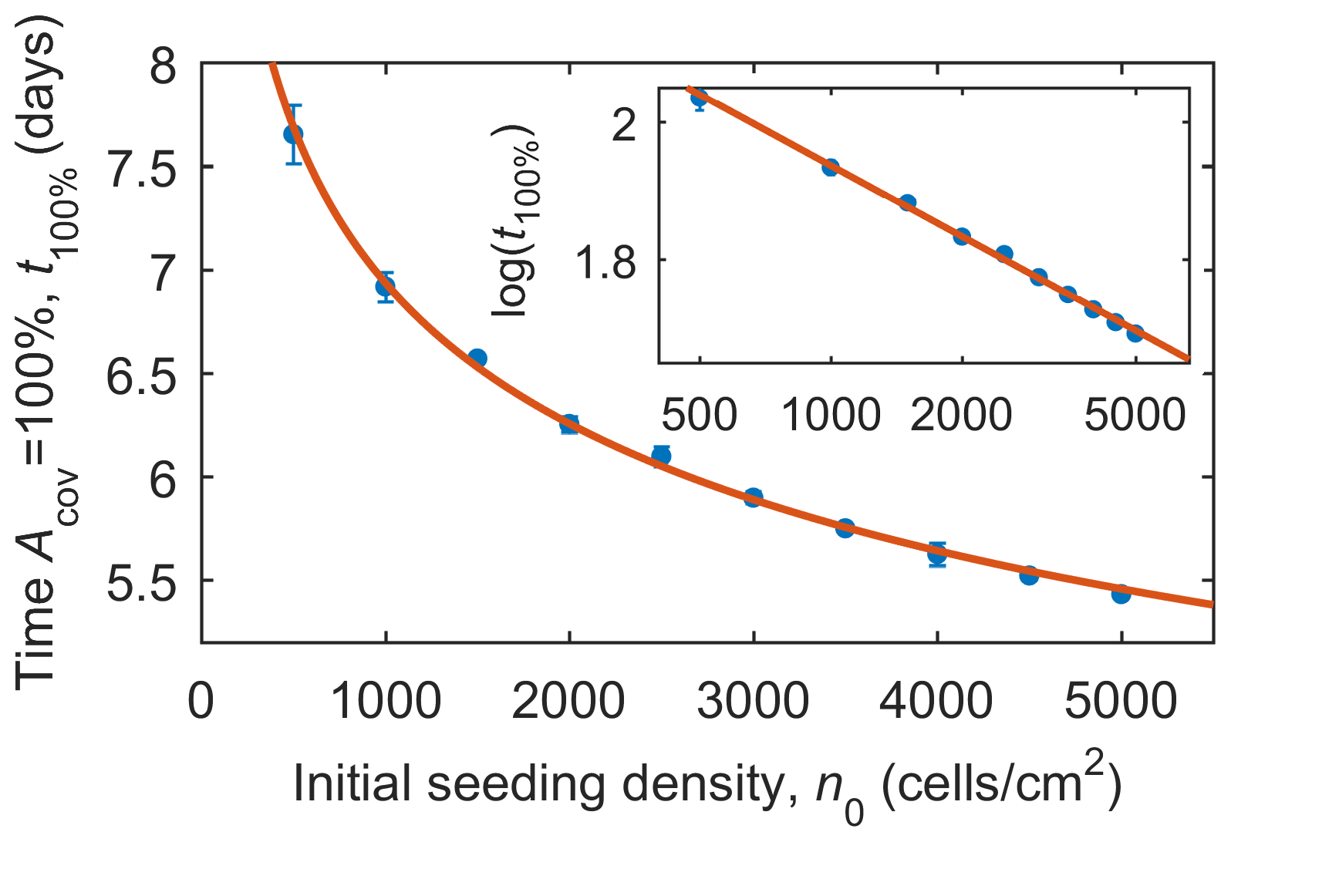}
\label{fig:coverage2}
\end{subfigure}
\caption{\label{fig:coverage} (\subref{fig:coverage1}) The percentage area coverage of cells at different time points for different cell seeding densities according to the two-population model. Inset: The data on a log-linear scale. (\subref{fig:coverage2}) The time the growth area is 100\% covered for varying cell seeding densities, with line of best fit $t_{\rm{100\%}}=20n_0^{-0.15}$. Inset: The data on a log-log scale, with line of best fit $\log(t_{\rm{100\%}})=(-0.15\pm0.005)\log(n_0)+(3.0\pm0.04)$.}
\end{figure}

Now the initial conditions of cell seeding are known, the growth of colonies from these cells can be considered. Cells are seeded at density $n_{\rm{0}}$ according to a Poisson point process as described above, and then the division of the cells and growth into colonies can be described by the two-population model. The area coverage of the plate can be estimated from the number of cells we expect to be present. The average area of a cell, $A_{\rm{cell}}$, is approximately 250\,$\mu$m$^2$ [S. Orozco-Fuentes, private communication], corresponding to a cell of size 18\,$\mu$m. The percentage of area covered by cells evolves as shown in Figure~\ref{fig:coverage1}. Taking this value, the proportion of area coverage is the area covered by the cells, $N(t)A_{\rm cell}$, divided by the growth area of the plate and $A_{\rm{plate}}=N_{\rm{seeded}}/n_0$. We therefore expect the percentage area coverage over time to tend to an exponential relationship due to the growth of $N(t)$, scaled by a factor equal to $n_0A_{\rm{cell}}/N_{\rm{seeded}}$, as we see in Figure~\ref{fig:coverage1}. The time taken for the growth area to be 100\% covered, $t_{\rm{100\%}}$, for varying initial seeding densities, is shown in Figure~\ref{fig:coverage2}. 

\begin{figure}[!]
\centering
\begin{subfigure}{0.2\textwidth}
\caption{}
\includegraphics[width=\textwidth]{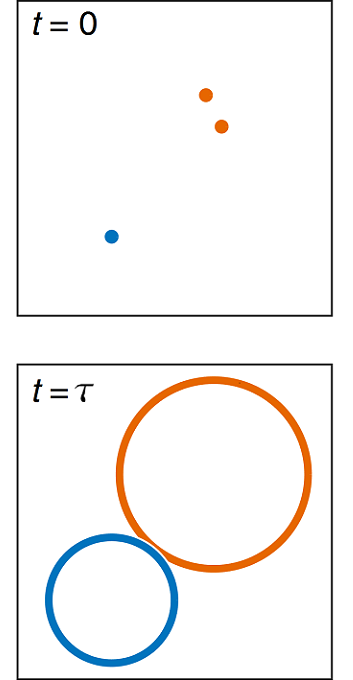}
\label{fig:tmerge1}
\end{subfigure}
\begin{subfigure}{0.255\textwidth}
\caption{}
\includegraphics[width=\textwidth]{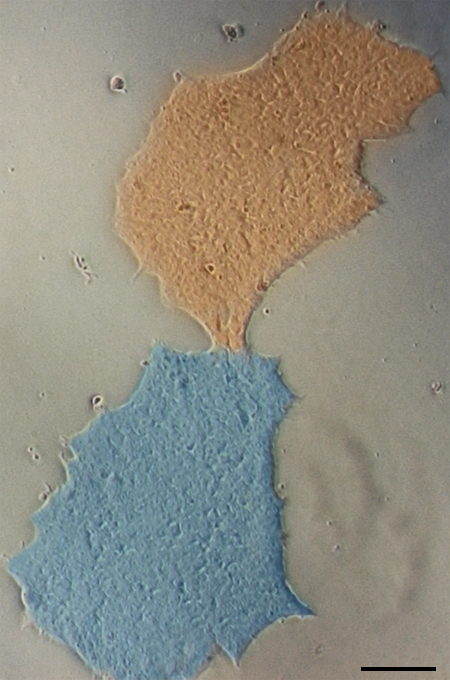}
\label{fig:tmerge2}
\end{subfigure}
\hspace{10pt}
\begin{subfigure}{0.49\textwidth}
\caption{}
\includegraphics[width=\textwidth]{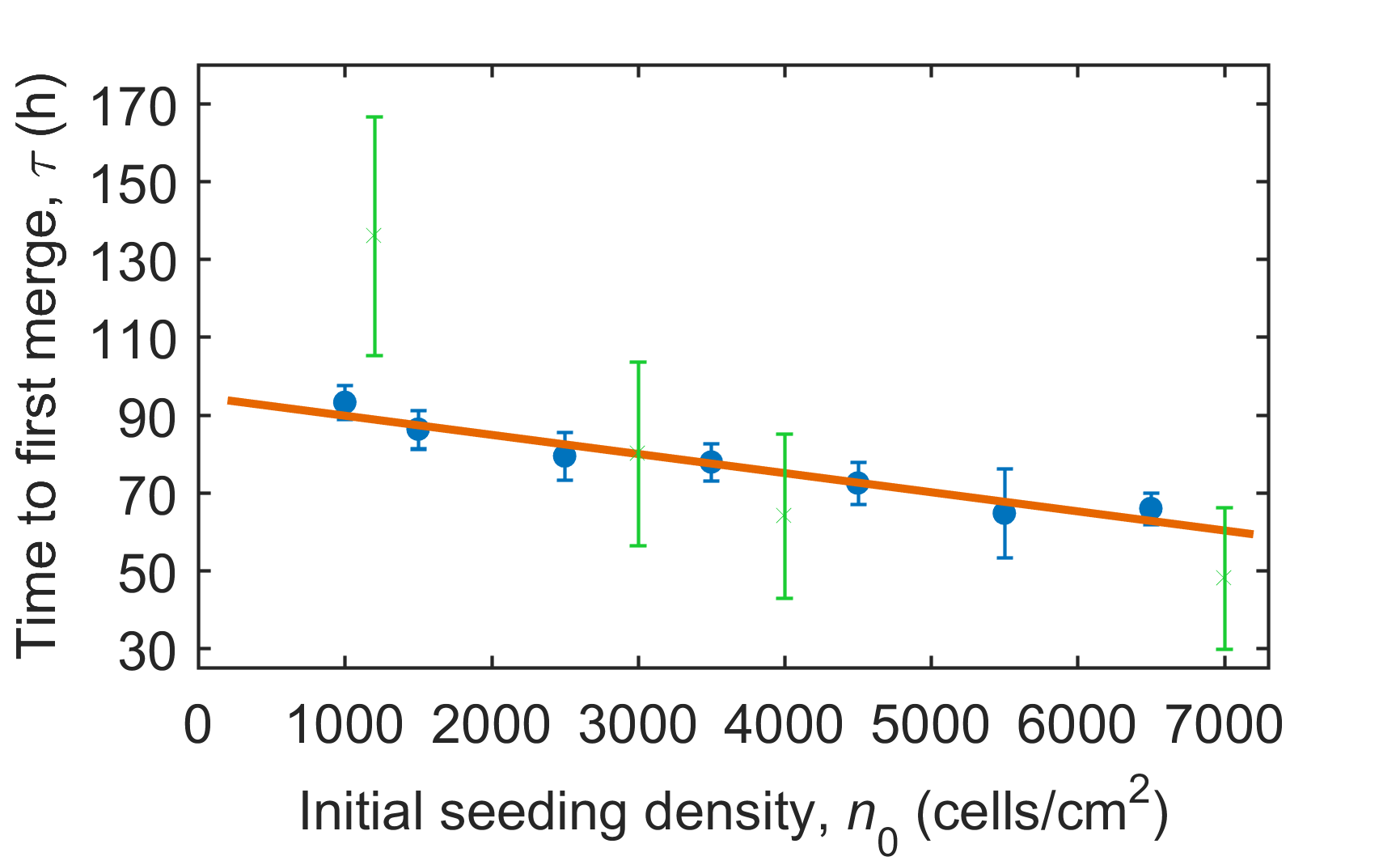}
\label{fig:tmerge3}
\end{subfigure}
\caption{\label{fig:tmergeall}(\subref{fig:tmerge1}) Diagram illustrating initially seeded cells, and $\tau$, the first time at which two growing colonies touch from a simulation of the cell seeding model. The orange cells are classed as a pair and grow accordingly. (\subref{fig:tmerge2}) An example of two colonies merging from experimental images. The two colonies, highlighted in blue and orange are beginning to merge together at 5 days. The scale bar represents 100$\,\mu$m. (\subref{fig:tmerge3}) The mean time (averaged over 20 simulations) that the first colony merge occurs for cells seeded at different densities growing according to the two-population model. The line of best fit, $\tau=(-0.005\pm0.004)n_{\rm{0}}+(95\pm6)$ with $R^2=0.93$ is shown in orange. The mean of three experimental values of $\tau$ are shown as green crosses with error bars calculated through error propagation based on an error of $\pm\,0.5$ days for each of the measured values due to the time resolution of the images.}
\end{figure}

Simulating the initial conditions as described above and the colony growth allows us to estimate the crucial time at which the colonies begin to merge. The cells are seeded at density $n_0$, with $\eta_0$ cells attached, and are then sorted into clusters based on their spatial distances away from each other. Each cluster grows according to the two-population model, estimated as a circle with centre at the geometric centre of the cluster and radius based on the number of cells present, $N(t)$. The growth rate for triples and larger clusters of cells is assumed to be the same as that for pairs of cells. The time at which any colony begins to merge with its neighbour is critical as the time that clonality is lost, $\tau$, illustrated in Figure~\ref{fig:tmerge1}. The time the first colony merge occurs at varying seeding densities is shown in Figure~\ref{fig:tmerge2}, with least squares fitting $\tau=(-0.005\pm0.004)n_{\rm{0}}+(95\pm6)$ with $R^2=0.93$, $\tau$ in hours and $n_0$ in cells/cm$^2$. We are therefore able to estimate the time taken for the first colony merge to occur from the equation 

\begin{equation}
 \frac{\tau}{1\,\textrm{h}} \approx 95-\frac{n_0}{200\,\textrm{cm}^{-2}},
\end{equation}

\noindent
where $n_0$ is the initial seeding density of cells before attachment in cells/cm$^2$ and $\tau$ is produced in hours. Experimental values were extracted for $\tau$ from Experiment 2 and the model captures these values within errors for the seeding densities 3000, 4000 and 7000\,cells/cm$^2$. These results are summarised in Table~\ref{tab:seeding} for convenience.

\begin{table}[ht]
\centering
\resizebox{\textwidth}{!}{\begin{tabular}{cccc}  
\toprule
Seeding density, $n_0$ (cells/cm$^2$) & Attached cells, $\eta_0$ (cells/cm$^2$) & Percentage of single cells & Time to first colony merge, $\tau$ (h)\\
\midrule
1000   & 350 $\pm$ 42 &  78 $\pm$ 3& 90 $\pm$  10   \\
1500   & 525 $\pm$ 63  & 70 $\pm$ 3& 87  $\pm$ 12    \\
2000   & 700 $\pm$ 86 & 62 $\pm$ 2& 84  $\pm$ 14  \\
3000   & 1050 $\pm$ 127 & 48 $\pm$ 1& 80 $\pm$ 18   \\
5000   & 1750 $\pm$ 212 & 30 $\pm$ 1& 70 $\pm$  26   \\
\bottomrule
\end{tabular}}
\caption{The expected number of attached cells, single cells at seeding and the time to the first colony merge for varying seeding densities.}\label{tab:seeding}
\end{table}

\section*{Discussion}

Colony growth originating from single or pairs of cells is well modelled using a two-population stochastic exponential model where the growth rate is sampled from a Normal distribution. Experimental results show that colonies that start from a single founder cell grow at a rate different from those originating from a cell pair, with a relative difference of $8\pm1.8\%$. The colonies originating from pairs of cells have a higher mean growth rate (although it is within standard deviation errors of the growth rate for single cell colonies), and the standard deviation of the growth rate for pairs is much lower, as seen in Figure~\ref{fig:gammadist}. It is likely that the growth rates are actually time dependent, so although we see a difference in the growth rates from single and paired founding state up to 72\,h, we expect this difference to reduce and eventually become negligible over time. Similarly, we expect that the difference in growth rates between triples and other larger groups of founding cells also decrease as the cluster size increases. Further experimental data would be needed to clarify this. This model can be used to predict colony sizes at different time scales. We thus suggest that single clone colonies (that originate from a single founder cell) can be selected as the smaller of the growing colonies, but only within a certain period, $t_{\ast}$, before colony sizes from a range of founder cells can be equivalent. This time gives an indication of when colonies should be observed to identify single founder cell colonies based on colony size. This time, $t_\ast$, ranges from around 30 to 50\,h depending on the seeding density.

Experimental results show that, on average, $35\%$ of initially seeded cells are attached to the plate 24 hours after cell seeding, Figure~\ref{fig:attachment}. We take this into account when considering the initial conditions of cell seeding. It is thought that cells affect each other's movement if they are less than 150\,$\mu$m apart \cite{Li}. The nearest neighbour distributions for cells seeding according to a Poisson process, shown in Figure~\ref{fig:nearestn}, indicates that for a low seeding density of $n_0=1500$\,cells/cm$^2$ (i.e. around 500 attached cells) we would expect to see around 30\% of cells with a nearest neighbour closer than the critical value $d=150\,\mu$m. However, our previous results suggest that we wouldn't expect all these pairs to migrate towards each other and form colonies \cite{me1}. The initial clustering of cells is an important and non-trivial consideration with implications on cell communications. We define a cluster to be a group of cells where each cell is 150\,$\mu$m or closer to another cell in the cluster. This allows clusters of different structures, as illustrated in Figure~\ref{fig:clusteringdiag}. This random seeding process gives the proportion of cell clusters of different sizes presented in Figure~\ref{fig:clusteringgraph}. At low seeding densities up to $\approx 3000$\,cells/cm$^2$, i.e. $\approx 1000$ attached cells, single cells with no neighbours dominate, making up over 50\% of cells present. As the seeding density increases the amount of single cells decreases, with the number of cells in pairs, triples and larger groups increasing (each proportion will always tend to zero as a wider variety of cluster sizes appear). This gives an indication of the seeding densities required to ensure a large proportion of the colonies originate from a single founder cell. 

Following the population growth model we can also estimate the time to completely fill a growth area at different seeding densities, Figure~\ref{fig:coverage}. A more interesting time to consider is the average time that growing colonies first merge due to physical proximity at different seeding densities, Figure~\ref{fig:tmerge2}. This time, $\tau$, has a linear line of best fit with $\tau\approx95-0.005n_0$, allowing us to predict this time for any seeding density. This statistical estimation of $\tau$ suggests the latest possible observation time to catch colonies for re-plating before they merge. These results can be used to inform cell seeding density choices to maximise clonal colonies and avoid those arising from more than one founder cell.

\section*{Methods}

Two types of experiment were carried out: Experiment 1 collected data on colony numbers at 72 hours and Experiment 2 collected data on the rates of cell attachment and the time to colony merging. 

Experiment 1: Human embryonic stem cells (WiCell, Madison WI) were plated at a density of 1500 cells/cm$^2$ onto 6-well plates coated with Matrigel\textsuperscript{\textregistered} Basement Membrane Matrix (Corning Inc.), in the mTeSR1\textsuperscript{TM} media (STEMCELL Technologies). The cells were stained with Cell Trace Violet Dye (Thermo Fisher). At 72\,hours after cell attachment the cells were fixed and microscopy images (Nikon Eclipse Ti-E microscope) were taken of the colonies. Data was then collected using Imaris Image Analysis Software (BITPLANE Inc) to identify cell boundaries and count the number of cells in each colony. This data was extracted  for 48 colonies. 

Experiment 2: Following the same set-up as Experiment 1, the numbers of cells attached 24 hours after seeding were recorded for different initial densities of 1000, 1200, 2000, 3000, 4000 and 7000 cells/cm$^2$. Microscopy images were also taken of these wells each day for eight days for the initial seeding densities 1200, 3000, 4000 and 7000\,cells/cm$^2$. The time the first colony merge occurred at each seeding density was extracted from examination of the images.

\section*{Data Availability}
The datasets generated during and analysed during the current study are available from the corresponding author on reasonable request.

\bibliography{mybib}

\section*{Acknowledgements}

We acknowledge financial support from Newcastle University and European Community (IMI-STEMBANCC, IMI-EBISC, ERC \#614620 and NC3R NC/CO16206/1) and are grateful to the School of Mathematics, Statistics and Physics of Newcastle University (Prof. R. Henderson) for providing partial financial support. SOF thanks the National Council for Science and Technology (CONACYT), Mexico, for the scholarship CVU-174695. ML acknowledges BBSRC UK (BB/I020209/1) for providing financial support for this work. AS acknowledges partial financial support of the Leverhulme Trust (Grant RPG-2014-427).

\section*{Author contributions statement}

IN, ML, NGP and AS designed and supervised the study. IN and SB conducted the experiments. LEW analysed the data, developed the modelling and prepared the figures. SOF contributed to group discussions. AL performed the imaging. LEW, SOF, IN, NGP, ML and AS wrote the manuscript. 

\section*{Additional information}

\textbf{Competing interests}: The authors declare that they have no competing interests. 

\noindent
\textbf{Supplementary information}:...

\end{document}